\documentclass[12pt,twoside]{article}
\usepackage{xrb2007}
\usepackage{graphicx}

\def\approxgt{\ifmmode \rlap{$>$}{}_{{}_{{}_{\textstyle\sim}}} \else%
$\rlap{$>$}{}_{{}_{{}_{\textstyle\sim}}}$\fi} 
\def\approxlt{\ifmmode \rlap{$<$}{}_{{}_{{}_{\textstyle\sim}}} \else%
$\rlap{$<$}{}_{{}_{{}_{\textstyle\sim}}}$\fi}

\def\flx{erg cm$^{-2}$ s$^{-1}$}

\def\xmm{XMM-{\it Newton}}
\def\chan{{\it Chandra}}

\begin{document}

% select your session by uncommenting the appropriate line
%\session{Jets}
%\session{Jet and Black Hole Binaries}
\session{Faint Galactic XRB Populations}
%\session{Faint XRBs and Galactic LMXBs}
%\session{Obscured XRBs and INTEGRAL Sources}
%\session{ULXs}
%\session{Extragalactic Populations}
%\session{Future Missions and Surveys}
%\session{Population Synthesis}

\shortauthor{Jonker, Torres, and Steeghs}
\shorttitle{Faint Galactic X--ray Binaries}

\title{Faint Galactic X--ray Binaries}
\author{P.G.~Jonker}
\affil{SRON, Netherlands Institute for Space Research, 3584~CA, Utrecht, The Netherlands\\Harvard--Smithsonian  Center for Astrophysics, Cambridge, MA~02138, Massachusetts, U.S.A.}
\author{M.A.P.~Torres}
\affil{Harvard--Smithsonian  Center for Astrophysics, Cambridge, MA~02138, Massachusetts, U.S.A.}
\author{D.~Steeghs}
\affil{Astronomy and Astrophysics, Department of Physics, University of Warwick, Coventry
CV4~7AL, U.K.\\Harvard--Smithsonian  Center for Astrophysics, Cambridge, MA~02138, Massachusetts, U.S.A.}

\begin{abstract}
We present a short overview of the properties of faint Galactic X--ray
binaries. We place emphasis on current classification scenarios. One
of the important parameters for the faint sources is their intrinsic
luminosity. In the case of low--mass X--ray binaries it has recently
been realised that besides a phase of radius expansion, the duration
of type I X--ray bursts can be used as a primer for the source
luminosity in some cases. Further, we show that a very low equivalent
width of hydrogen and helium emission lines in the optical spectrum
alone is not a tell--tale sign for an ultra--compact system. Finally,
we list and discuss some unusual sources that could be X--ray
binaries.

\end{abstract}

\section{Introduction}

In recent years many new and often faint X--ray binaries have been
discovered.  Sources are found in the deep images made in hard X--rays
by INTEGRAL and in images obtained by {\it Swift}, \xmm\ and
\chan. The sensitivity to soft X--rays of the CCD instruments on board
the latter satellites as well as their small point--spread function
allows for optical or near--infrared identification of the counterpart
to the X--ray source which is essential for source
classification. Especially for sources in the Galactic plane, crowding
often requires the superb \chan\ positional accuracy and precludes the
use of the existing optical and near--infrared sky survey data such as
that of the (S)DSS and 2MASS. Although, in the case of obscured
high--mass X--ray binaries as well as the supergiant fast X--ray
transients, lower resolution X--ray images together with especially
2MASS often provide a fair assessment of the source type. There are
currently several programs providing identifications of the newly
discovered sources (e.g.~see various contributions to these
proceedings and \citealt{2005ATel..494....1S},
\citealt{2005ATel..629....1S}, \citealt{2007ATel.1072....1T},
\citealt{2007ATel.1193....1T}, \citealt{2006ApJ...647.1309T},
\citealt{2008arXiv0802.0988M}, \citealt{2008arXiv0802.1774C} to cite
just a few).

\section{Identification}

\begin{figure} 
\centering
\includegraphics[scale=.5,angle=270]{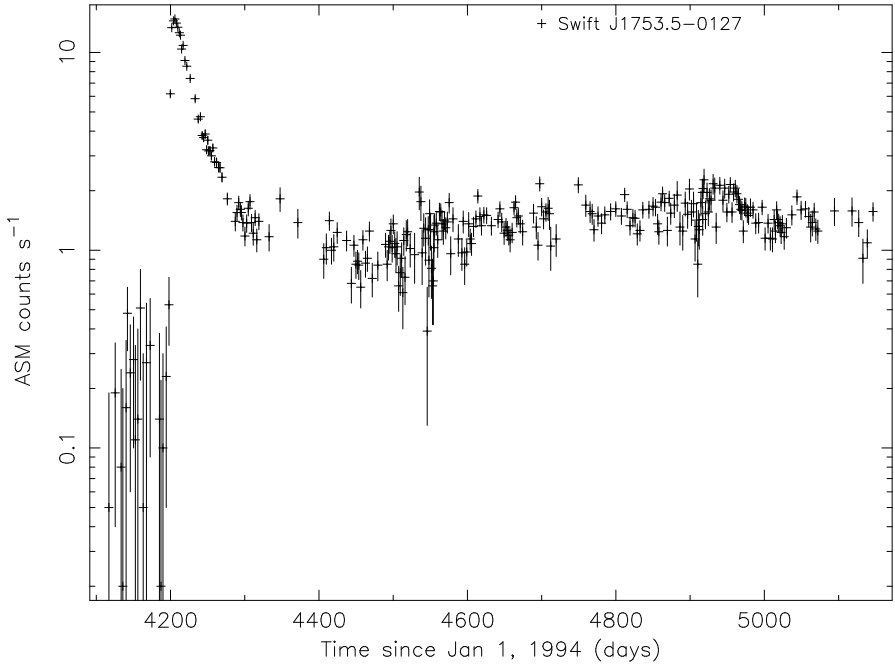} 
\caption{Light curve of the black hole candidate Swift~J1753.5--0127
obtained by the All Sky Monitor onboard the Rossi X--ray timing
Explorer. The source is a long--duration transient where the outburst
lasts several years (see also
\citealt{2007ApJ...659..549C}). \label{lc1753}}
\end{figure}

The increase of the sample of sources has also enlarged the covered
parameter space. New sources sometimes display rare phenomena and
properties that were thought to be unique for a certain class turned
out to be more common. For instance, more and more faint transient and
faint persistent low--mass X--ray binaries have been identified
(\citealt{2003ApJ...598..474M}, \citealt{2006MNRAS.366L..31K},
\citealt{2007A&A...465..953I}, and \citealt{2008arXiv0801.0953W} these
proceedings). In addition, there have been several new long--duration
transients (e.g.~the transient Z source XTE~J1701--462;
\citealt{2007ApJ...656..420H}), of which two sources still remain
active today (the black hole candidate SWIFT~J1753.5--0127, see
Figure~\ref{lc1753}, and the accretion powered ms X--ray pulsar
HETE~J1900.1--2455; \citealt{2007ApJ...654L..73G}). Two new groups of
high--mass X--ray binaries have recently been identified. The obscured
high--mass X--ray binaries (e.g.~IGR~J16318--4848;
\citealt{2003A&A...411L.427W}) and the supergiant fast X--ray binaries
(e.g.~\citealt{2005A&A...444..221S}, \citealt{2007arXiv0704.3224N}).

Many new X--ray sources have been discovered in the \chan\ survey
described in \citet{2002Natur.415..148W} and
\citet{2003ApJ...589..225M}. Considerable effort has gone into the
source classification but the large extinction and source crowding in
fields near the Galactic Centre hamper the identifications
(\citealt{2005MNRAS.364.1195B}, but see \citealt{2006SPIE.6269E..39E}
and Eikenberry et al.~these proceedings). Considerable effort is also
put--in in the ChamPlane survey (\citealt{2005ApJ...635..920G},
\citealt{2005ApJ...634L..53L}).

As was indicated in a recent paper by \citet{2008arXiv0801.1101M},
some of the faint X--ray transients found towards the Galactic Centre
might be dwarf Novae in outburst. In the absence of the tell--tale
signs of a neutron star (pulsations, type~I X--ray bursts) a good way
to distinguish between a nova or a low--mass X--ray binary origin is to
look at the ratio of the X--ray and optical luminosity and at the
equivalent width of the hydrogen and helium emission lines in the
optical spectra. One can determine the equivalent width of the
emission lines that are typically found in interacting and active
binaries and AGN even with medium resolution spectra. The equivalent
width, the ratio of optical and X--ray flux (and hence optical and
X--ray luminosity), and the estimate of the optical magnitude of the
source allow one to classify the sources. For instance, for a given
optical to X--ray flux ratio the equivalent width of H and He emission
lines in cataclysmic variables is much larger than that of low--mass
X--ray binaries (\citealt{1985ApJ...292..535P},
\citealt{1995xrb..book...58V}).

High--mass X--ray binaries can be separated from cataclysmic variables
and low--mass X--ray binaries by the fact that their optical and/or
near--infrared counterparts are several magnitudes brighter. Even when
extinction is high and the distance large a near--infrared ($K$--band)
image will reveal a bright counterpart if the source is a high--mass
X--ray binary (e.g.~\citealt{2003A&A...411L.427W},
\citealt{2006ATel..831....1N},
\citealt{2006ApJ...647.1309T}). Finally, often excess extinction over
the Galactic interstellar value is found, implying that the extinction
is intrinsic to the source.  A large amount of local extinction points
to a high--mass rather than a low--mass system. However, care should
be taken as variations in the extinction can be large and on much
smaller scales than provided by the \citet{1990ARA&A..28..215D} or
\citet{scfida1998} maps (e.g.~see the work of
\citealt{2006ApJ...640L.171G}).

\section{The Luminosity of Faint Sources}

One of the important unknown parameters of the faint persistent and
transient sources is their intrinsic luminosity. Do these sources
appear faint due to e.g.~their large distances, a large interstellar
extinction, or due to inclination effects, or are they intrinsically
weak? It is well known that inclination effects can be important. In
high inclination systems the accretion disk can occult part of the
inner disk from our view which can lead to a lower apparent
luminosity. Similarly, (mild) beaming effects
(\citealt{2002MNRAS.335L..13K}) and scattering could be important
(\citealt{1978A&A....67L..25M}, \citealt{1985MNRAS.217..291L},
\citealt{1988ApJ...324..995F},
\citealt{2005ApJ...623.1017N}). Nevertheless, inclination and beaming
effects are not a likely explanation for the faintness of all the
faint sources that are found.

An existing method to determine distances in neutron star low--mass
X--ray binaries relies on the occurrence of photospheric radius
expansion thermonuclear explosions on the surface of low magnetic
field accreting neutron stars (so called type I X--ray bursts, for
a calibration of the distance scale, theoretical background and
applications see \citealt{2003A&A...399..663K},
\citealt{2003ApJ...590..999G,2006ApJ...639.1033G},
\citealt{2004MNRAS.354..355J}). In order to determine whether the
photosphere undergoes radius expansion one needs to obtain X--ray data
with at least modest energy resolution. However, both recent
observational and theoretical work on these type~I X--ray bursts show
that they can be used to infer whether a source is intrinsically faint
or not without the need for photospheric radius expansion bursts
(\citealt{2007ApJ...654.1022P}, \citealt{2007A&A...465..953I}). If the
burst duration of a type~I X--ray burst is longer than a few minutes
it implies that a rather thick layer of accreted material has been
burned in the thermonuclear explosion. The conditions to build up such
a thick layer require a luminosity below a few percent of
Eddington. \citet{2007A&A...465..953I} applied this method, as well as
the distance derived from radius expansion bursts (when present), and
concluded that the intrinsic luminosity of a sample of faint
persistent sources is low. At such low luminosities it is difficult
for the accretion disk to be fully ionized, hence these systems should
be transients unless they are ultra--compact X--ray binaries with
orbital periods less than $\sim$1 hr.

\begin{figure} 
\centering
\includegraphics[scale=.5,angle=270]{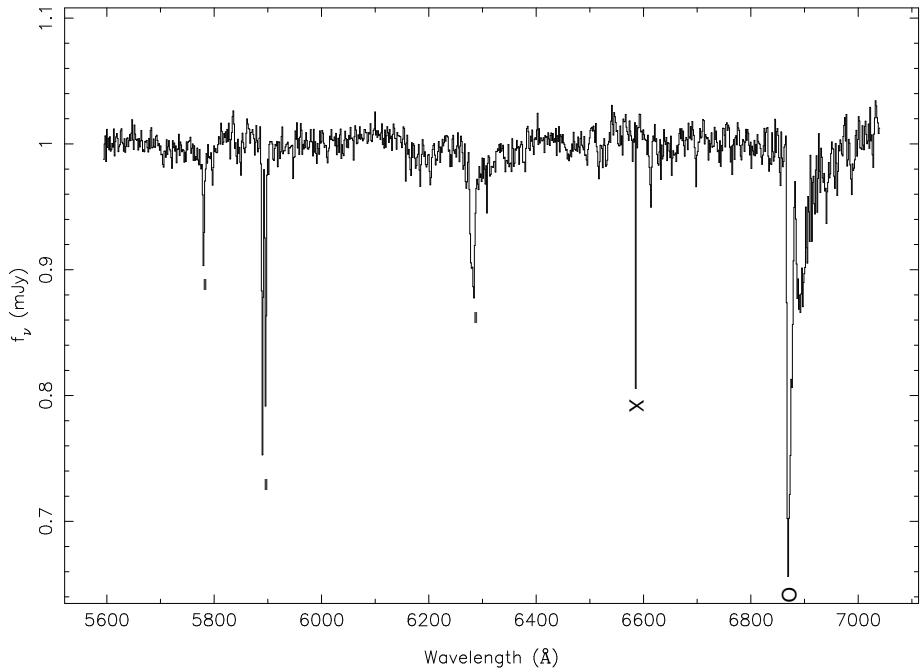} 
\caption{Optical spectrum of the counterpart of Swift~J1753.5--0127
obtained with the LDSS--3 spectrograph mounted on the 6.5~m Magellan
Clay telescope on June 25th 2006 1:45-5:15UT. This spectrum is the
average of 13$\times$600 s exposures. Observing conditions were good
with seeing around 0.9-1.0". The VPH-Blue grism was employed with a
0.75" slit. The small vertical bars indicate absorption lines caused
by interstellar absorption. The cross indicates a spike that is caused
by a CCD defect and the open circle indicates a feature caused by
telluric absorption.
\label{j1753sp}}
\end{figure}

In optical spectroscopic observations obtained with the VLT of one of
the sources studied by \citet{2007A&A...465..953I}, 1A~1246--599
(\citealt{2006A&A...446L..17B}, In 't Zand et al.~submitted), the
hydrogen and helium emission lines normally found in low--mass X--ray
binaries are not detected down to equivalent width limits that are
much lower than those observed in low--mass X--ray binaries that have
orbital periods in excess of a few hrs (see also
\citealt{2005AIPC..797..396N}, \citealt{2006MNRAS.370..255N}). This,
together with the low optical brightness of the source strongly argues
for an ultra--compact X--ray binary nature of this
source. Interestingly, in an optical spectrum of the transient black
hole candidate Swift~J1753.5--0127 we also did not find evidence for
hydrogen or helium emission lines (see Figure~\ref{j1753sp},
\citealt{2007ApJ...659..549C}), however, the orbital period of this
system is 3.2 hours (\citealt{2007ATel.1130....1Z}). Therefore, the
absence of hydrogen and or helium emission lines alone cannot be used
as evidence for an ultra--compact nature of the source. Furthermore,
earlier in the outburst optical spectra of Swift~J1753.5--0127 did
show H$\alpha$ emission (\citealt{2005ATel..551....1T},
\citealt{2007ApJ...659..549C}).

In the case of high--mass X--ray binaries the distance can be
estimated in a crude way by using the approximate location of spiral
arms and the Galactic position of the source. These relatively young
and massive systems reside close to spiral arms.  Some sources appear
faint due to the fact that the interstellar extinction causes a
reduction in the observed amount of soft X--rays. However, even though
many of the faint obscured sources discovered by INTEGRAL appear faint
in the soft X--ray bands due to the large extinction, the extinction
is much less important above a few keV and many of these sources have
luminosities above 1$\times 10^{35}$ \flx
(e.g.~\citealt{2003A&A...411L.427W},
\citealt{2006A&A...448.1101I}). Hence, observing at high X--ray
energies reduces/eliminates the effects of extinction.

\section{Outliers; Rare Transients}

The diversity among the properties of X--ray binary transients is
large.  Nevertheless, recently a handful of transients has been
described that do not fit the known categories. In Table 1 we list the
ones currently known to us.

\begin{table}
\caption{Rare transients}
\begin{tabular}{lcc}
\hline
Source name & Peak flux & Reference \\
\hline
Swift~J1749.4-2807 &  2$\times 10^{-8}$ \flx (15--50 keV) & Wijnands et al.~2007\\
Swift~J195509.6+261406	& 5$\times 10^{-8}$ \flx (15--150 keV)	& Kasliwal et al.~2007\\
XTE~J1901+014$^*$	& 0.5-1 Crab & \citealt{2007AstL...33..159K}\\
IGR~J11321-5311 & 2.2$\times 10^{-9}$ \flx (20--300 keV) & \citealt{2007AA...468L..21S}\\
Ginga Burst 900129	& --	& \citealt{1995ApJ...445..731S}\\
\end{tabular}
{\footnotesize$^*$ Possible fast supergiant transient?}\\
\end{table}

\citet{2007arXiv0709.0061W} classify Swift~J1749.4-2807 as a fast
transient low--mass X--ray binary at a distance of 6.7$\pm$1.3
kpc. They interpret the Swift/BAT trigger as due to a type I X--ray
burst. The evidence they provide to support their claim comes from the
\xmm\ discovery of a point source at the position of the Swift/XRT
afterglow both 6 years before the burst as well as 4 months after the
burst (\citealt{2007arXiv0709.0061W}). The Swift/XRT afterglow
emission decreased at a huge rate (a factor 1000 in less than a
day). This observation increases the parameter space of the decay rate
of low--mass X--ray binary transients enormously (the black hole
transient XTE~J1908+094 showed a decay rate of a factor of 750 in less
than 24 days; \citealt{2004MNRAS.inpress}).

Like Swift~J1749.4-2807, Swift~J195509.6+261406 was first reported to
be a Gamma--ray Burst (GRB; \citealt{2007GCN..6489....1P}). However,
the subsequent Swift/XRT and near--infrared monitoring
(\citealt{2007arXiv0708.0226K}) showed that the afterglow was unlike
that of a GRB. Kasliwal et al.~(2007) propose the source to be a black
hole X--ray transient. If indeed low--mass X--ray binaries are
responsible for the two Swift events noted above they can by extension
also be responsible for the other events listed in Table 1. An
alternative interpretation for some of the events in Table 1 is that
they are caused by a Soft Gamma--ray Repeater or an Anomalous X--ray
Pulsar (SGR, AXP; \citealt{2006csxs.book..547W}).

\begin{figure} \includegraphics[scale=0.95,angle=0]{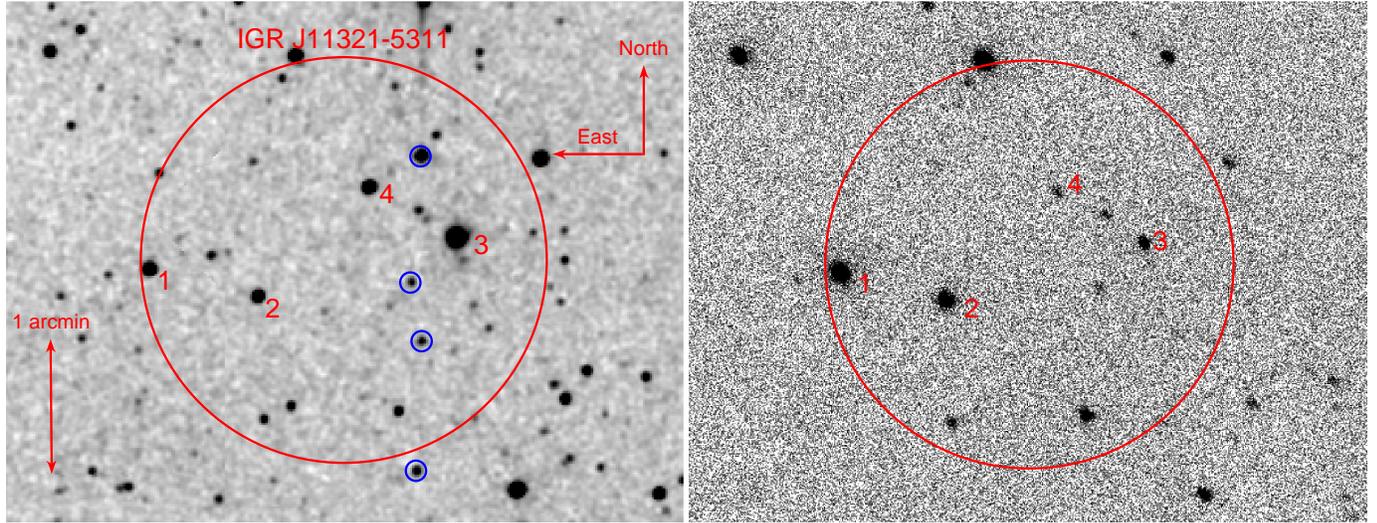}
\caption{{\it Left panel:} Near--infrared 2MASS image ($K$--band) of
the field of IGR~J11321-5311. {\it Right panel:} Swift/UVOT
$UVW2$--band image of the same area, showing that the stars labelled 1
through 4 are in the foreground of the dark cloud that is centered
near star 3 and encompasses the whole field of view displayed here
(\citealt{1984A&AS...58..365F}; \citealt{1994JApA...15..415R}). The
four encircled stars are not listed in the 2MASS catalogue, even
though they are significantly detected. \label{2mass1132}}
\end{figure}

Finally, we repeat that some of the (obscured) fast hard X--ray
transients that have been discovered in recent years also exhibit
large fast flares (\citealt{2006ApJ...638..974S}), hence such sources
could be responsible for some of the transient events reported above,
most notably XTE~J1901+014 and IGR~J11321--5311. However, the hard
X--ray spectrum of IGR~J11321--5311 shows an upturn at the lowest
energies that are covered by the INTEGRAL/ISGRI instrument which is
not found in the other fast hard X--ray transients. Furthermore, for
most of the supergiant fast X--ray transients multiple outbursts have
been detected
(e.g.~\citealt{2005A&A...444..221S,2007arXiv0704.3224N}). So far, that
is not the case for the systems listed in Table 1. Finally, there is
no strong candidate for a supergiant counterpart in the 2MASS
catalogue (\citealt{2006AJ....131.1163S}) in the INTEGRAL error circle
of IGR~J11321--5311 (see Figure~\ref{2mass1132}). In the direction of
the transient there is a dark cloud encompassing the full area shown
in Figure~\ref{2mass1132} (the dark cloud is centered on $l$=291.10,
$b$=7.85 which is close to star 3 in Figure~\ref{2mass1132};
\citealt{1984A&AS...58..365F}; \citealt{1994JApA...15..415R}). Due to
the probably close proximity of the cloud ($<$400 pc;
\citealt{1994JApA...15..415R}), it is likely that the hard X--ray
transient is located behind this cloud. Since the extinction caused by
the dark cloud will cut out the UV light all the objects detectable
in the Swift $UVW2$ image should be foreground objects. If so, then
there are no bright 2MASS objects located in the INTEGRAL error circle
and behind the dark cloud that suggest a supergiant. Note that the
sources encircled in the {\it left } panel in Figure ~\ref{2mass1132}
have not been identified in the 2MASS catalogue. Due to the fact that
these sources all lie along the direction of the trail caused by a
very bright star just off the top the displayed field, we speculate
that this is due to the presence of this bright star
(2MASS~J11321193--5309324; K=5.771). Since this star is only just
outside the 90\% confidence error circle we cannot exclude that this
source is the counterpart to IGR~J11321--5311. Note, however, that it
is also detected in the UVOT image.

The deep limits on the near--infrared counterpart in quiescence of
Swift~J195509.6+261406 and the known interstellar extinction do
exclude the possibility of a early type star as the counterpart to
Swift~J195509.6+261406 (see Kasliwal et al.~2007).

\acknowledgements PGJ acknowledges support from the Netherlands
Organisation for Scientific Research. PGJ would like to thank
L.~Kuiper for useful discussions. DS acknowledges an STFC Advanced
Fellowship. This publication makes use of data products from the Two
Micron All Sky Survey, which is a joint project of the University of
Massachusetts and the Infrared Processing and Analysis
Center/California Institute of Technology, funded by the National
Aeronautics and Space Administration and the National Science
Foundation.

%\bibliographystyle{apj}
%\bibliography{0029-bib.bib}

\end{document}